\documentclass[twocolumn,aps,superscriptaddress,nofootinbib]{revtex4-1}
\usepackage{latexsym,amssymb,amsmath,epsfig,bm,psfrag}
\usepackage{color}
\usepackage[bookmarksnumbered,bookmarksopen,colorlinks,citecolor=red,linkcolor=blue]{hyperref}
\usepackage{mathrsfs}
\usepackage{times}
\usepackage{ulem}

\hypersetup{
    colorlinks=true,
    linkcolor=blue,
    filecolor=magenta,
    urlcolor=cyan,
    pdftitle={Sharelatex Example},
    bookmarks=true,
}

\newcommand\de{\delta e}
\newcommand\vk{{\bf k}}

\newcommand{\fig}[1]{Fig.~\ref{#1}}
\newcommand{\eq}[1]{Eq.~(\ref{#1})}

\newcommand\be{\begin{equation}}
\newcommand\ee{\end{equation}}
\newcommand\bra{\langle}
\newcommand\ket{\rangle}

\def\mfp{\lambda_{\rm mfp}}

\begin{document}

\title{Hydrodynamic attractor of noisy plasmas}
\author{Zenan Chen}
\email{19110200014@fudan.edu.cn}
\affiliation{Institute of Modern Physics, Fudan University, Shanghai 200433, China}

\author{Derek Teaney}
\email{derek.teaney@stonybrook.edu}
\affiliation{Department of Physics and Astronomy, Stony Brook University, Stony Brook, New York 11794, USA}

\author{Li Yan}\email{cliyan@fudan.edu.cn}
\affiliation{Institute of Modern Physics, Fudan University, Shanghai 200433, China}
\affiliation{Key Laboratory of Nuclear Physics and Ion-beam Application (MOE), Fudan University, Shanghai 200433, China}

\begin{abstract}

We provide a generalized formulation of fluctuating hydrodynamics for the far-from-equilibrium noisy medium. As an example, we consider a noisy plasma experiencing Bjorken expansion, for which the leading order evolution is captured by hydrodynamic attractor of classical hydrodynamics, while the quadratic couplings of fluctuations are solved effectively via a generalized version of the hydrodynamic kinetic equation. In the far-from-equilibrium plasma, backreaction of hydrodynamic fluctuations results in renormalization of transport properties, as well as long time tails, of high orders. In particular, corresponding to a renormalized hydrodynamic attractor, evolution in a noisy plasma towards equilibrium becomes non-monotonic. 

\end{abstract}
\maketitle

{\it Introduction.}---Hydrodynamics is an effective theory that by construction applies to thermal systems close to local equilibiurm. In hydrodynamics, 
departures from ideal fluids  
are captured by gradients of hydrodynamic fields 
as well as hydrodynamic fluctuations. 
Hydrodynamic fluctuations are in general suppressed in systems with large amount of constituents, it is therefore not surprising theoretical formulations without hydrodynamic fluctuations (frameworks sometimes referred to as the {\it classical} hydrodynamics \cite{Kovtun:2012rj}) have achieved remarkable successes. Such examples include in particular the 
hydrodynamic modeling of quark-gluon plasma (QGP) in high-energy nuclear physics~\cite{Shen:2020mgh}. 

Nonetheless, hydrodynamic fluctuations cannot be neglected when they are substantial to  
system dynamical evolution. 
For instance, when a thermal system evolves close to a critical point, correlations among fluctuations of order parameters diverge, resulting in novel hydrodynamic modes~\cite{Stephanov:2017ghc}. 
Hydrodynamic fluctuations are amplified in small systems, such as the QGP droplet created in high-energy proton-lead collisions~\cite{CMS:2015yux}, owing to the fact that correlations among thermal fluctuations are inversely proportional to the system volume. 
More importantly, the non-linear nature of hydrodynamics allows for corrections from couplings of fluctuations~\cite{POMEAU197563}. Backreaction of the coupled modes renormalizes transport properties~\cite{Kovtun:2011np,Akamatsu:2016llw,Akamatsu:2017rdu,Martinez:2017jjf,PhysRevD.84.025006}, generates non-analytical long time tail structures~\cite{Kovtun:2003vj}, and even influences evolution history in a fluid~\cite{Floerchinger:2014jsa}.

The equation of motion of fluctuating hydrodynamics follows the conservation of energy and momentum
\be
\label{eq:eom}
\partial_\mu T^{\mu\nu} = 0\,,\qquad
T^{\mu\nu} = T_{\rm cl}^{\mu\nu} + \delta T^{\mu\nu} + S^{\mu\nu}\,,
\ee 
where the classical energy-momentum tensor, $T_{\rm cl}^{\mu\nu}$, consists of energy density $e$, pressure $P$, fluid four-velocity $u^\mu$, and expansion in terms of their gradients~\cite{ff:footnote}, 
\begin{align}
\label{eq:cons}
T_{\rm cl}^{\mu\nu} = e u^\mu u^\nu + P \Delta^{\mu\nu} + \pi^{\mu\nu} \,.
\end{align}
The expansion can be characterized by the Knudsen number Kn.
Up to second order in gradients, the constitutive equation is often formulated via the stress tensor $\pi^{\mu\nu}$  
relaxing to its Navier-Stokes correspondence, 
i.e., the Israel-Stewart hydrodynamics~\cite{Israel:1979wp}. 
In the classical constitutive equation \eq{eq:cons}, 
variables are thermal averaged quantities without corrections from thermal fluctuations, namely, they are {\it bare} variables to be distinguished later from the renormalized ones. 
Fluctuations of energy-momentum tensor $\delta T^{\mu\nu}$ are constructed accordingly in terms of thermal fluctuations of hydrodynamic variables, which are further induced 
through the random noise $S^{\mu\nu}$, subject to the condition $\bra S^{\mu\nu}(x) \ket =0$ and the fluctuation-dissipation relation, 
$\bra S^{\mu\nu}(x)S^{\alpha\beta}(y) \ket = 2 T\eta\Delta^{\mu\nu\alpha\beta}\delta^4(x-y)$. Here, the brackets indicate an average over ensemble of thermal fluctuations. 
 With an equation of state: $P=P(e)$, Eqs.~(\ref{eq:eom}) and (\ref{eq:cons}) are closed.   

\eq{eq:eom} is stochastic hence the resulted system evolution fluctuates in space and time. However, the averaged evolution is deterministic, which can be obtained, for instance, through an ensemble average over 
numerical simulations of the stochastic processes. Alternatively, with respect to an effective field theory approach of fluctuating hydrodynamics~\cite{Kovtun:2014hpa,Liu:2018kfw,Crossley:2015evo}, by treating thermal fluctuations as perturbations, averaged quantities can be solved 
order by order. This is a strategy analogous to the loop expansion in quantum field theory. 
The tree-level analysis corresponds to 
solving the classical hydrodynamics, $\partial_\mu T_{\rm cl}^{\mu\nu}=0$. Effect of thermal fluctuations then arises 
when hydrodynamic fluctuations are included and constrained by $\partial_\mu \delta T^{\mu\nu} = - \partial_\mu S^{\mu\nu}$, which accordingly determines multi-point correlations~\cite{Akamatsu:2016llw,Akamatsu:2017rdu,An:2019osr,An:2020vri}. 
The two-point correlations, $\bra \delta T^{\mu\nu} \delta T^{\alpha\beta} \ket$, in particular, contain already the information of quadratic couplings of modes that contributes 
to $\bra T^{\mu\nu} \ket$, and
the renormalization of transport properties and the long time tails.

\begin{figure}
\begin{center}
\includegraphics[width=0.45\textwidth] {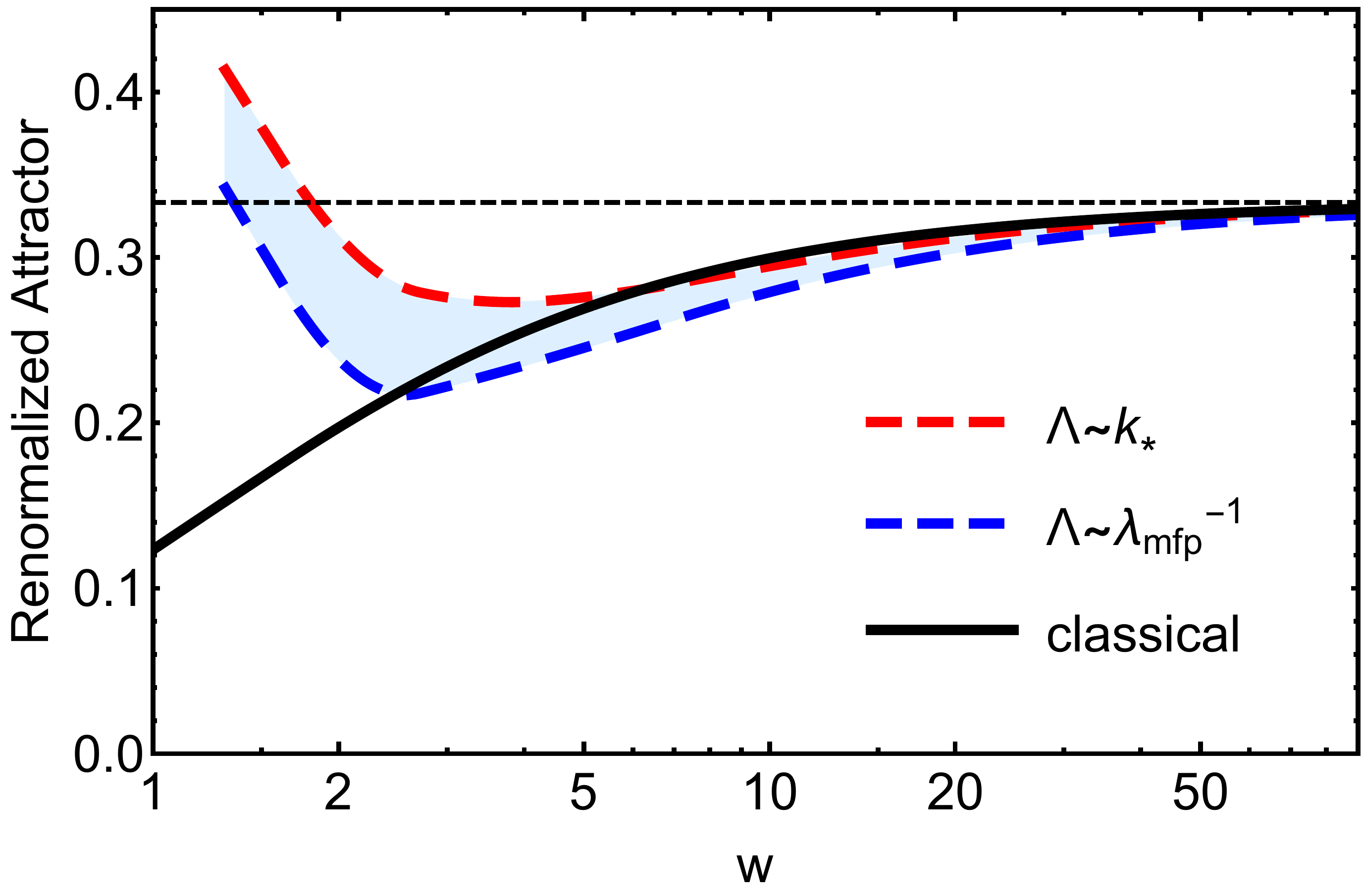}
\caption{
\label{fig:renAtt}
Isotropization of medium corresponding to attractors in classical hydrodynamics (black solid line) and fluctuating hydrodynamics (colored band). 
}
\end{center}
\end{figure}

Recently, extensive studies have been devoted to the generalization of {\it classical} hydrodynamics to far-from-equilibrium systems. These works are motivated in part by exploring the applicability condition of hydrodynamics 
through the convergence of gradient expansion~\cite{Heller_2013,Heller:2015dha}, and in part by the experimental observations of collectivity in QGP created from colliding nuclei of small sizes (cf. Ref.~\cite{CMS:2015yux}). From either aspect, it was acknowledged that classical hydrodynamics admits the so-called attractor solutions in some comoving flows~\cite{Romatschke:2017vte,Blaizot:2017ucy,Kurkela:2019set,Behtash:2017wqg,Denicol:2018pak,Jaiswal:2019cju,Chattopadhyay:2019jqj,Du:2021fok,Strickland:2018ayk,Brewer:2019oha,Behtash:2019txb}, 
owing to the expected hydrodynamic fixed points in these systems~\cite{Blaizot:2019scw}.

Bjorken flow, for instance, applies approximately to the very early stages of high-energy heavy-ion collisions, where system expands dominantly along the beam axis ($z$-axis). 
In the Milne coordinates, $\tau=\sqrt{t^2-z^2}$ and $\zeta=\tanh^{-1}(z/t)$, 
with respect to the  Israel-Stewart formulation, {\it classical} hydrodynamics reduces to coupled equations,
\begin{subequations}
\label{eq:eomBjork}
\begin{align}
\frac{d e}{d\tau} & = - \frac{1}{\tau}(e + P + \pi) \,,\\
\pi &=-\frac{4}{3}\frac{\eta}{\tau} - \tau_\pi \left[\frac{d \pi}{d \tau} + 
\frac{4}{3}\frac{\pi}{\tau}\right] \,, 
\end{align}
\end{subequations}
where $\pi=\pi^\zeta_{\;\zeta}$ is the $\zeta\zeta$-component of the stress tensor. For later convenience, we introduce dimensionless constants,
\be
\eta = C_\eta s\,, \quad
\tau_\pi = C_\tau C_\eta/T\,, \quad 
e = C_e T^4\,,
\ee
to parameterize shear viscosity $\eta$, shear relaxation time $\tau_\pi$, and energy density. 
We also consider the system conformal, which has $P=c_s^2 e$.

Defining the inverse Knudsen number Kn$^{-1}=w\equiv\tau/\tau_\pi$, Bjorken expansion of QGP is fully captured  by
the relative decay rate of energy density: $g(w)\equiv d\ln e/d\ln \tau$. Especially, isotropization of the system is related to $g(w)$ through
\be
\frac{P_L}{e}=
\frac{\tau^2 T_{\rm cl}^{\zeta\zeta}}{T_{\rm cl}^{\tau\tau}} = -1 - g(w)\,.
\ee
As will be clear later, this relation gets renormalized by hydrodynamic fluctuations.
In terms of $g(w)$, hydrodynamic attractor behaves as a smooth and monotonic connection between the free streaming fixed point at early times, $g(w)\approx -1$, and ideal hydrodynamic fixed point at late times, $g(w)= -4/3$, while evolution with arbitrary initial conditions merge swiftly towards the attractor.
Hydrodynamic attractor can be solved numerically, as well as analytically upon approximations~\cite{Blaizot:2020gql}. In the leading order adiabatic approximation~\cite{Blaizot:2021cdv} (or the leading order slow-roll approximation~\cite{PhysRevD.50.7222}), hydrodynamic attractor can be written as,
\be
\label{eq:gcl}
g(w) = -\frac{1}{2} \left[\frac{22}{7} + w  - \sqrt{\left(\frac{10}{21} + w \right)^2 + \frac{64}{45}}\right]\,.
\ee  
In the region $w\gtrsim 1$, \eq{eq:gcl} is not sensitive to second order transport coefficients~\cite{Blaizot:2021cdv}.

In \fig{fig:renAtt}, the istropization of medium corresponding to the {\it classical} hydrodynamic attractor,
\eq{eq:gcl}, is shown as the black solid line, which evolves monotonically from free streaming towards the ideal hydrodynamic fixed point, $1/3$, at late times. Note in particular, 
deviations from 1/3, at late times, are proportional to the bare shear viscosity $\eta$. 

Hydrodynamic attractor conceptually extends the applicability of classical hydrodynamics to systems with large Kn. It is not only because attractor universally describes system evolution irrespective of initial conditions, but also a consequence that attractor accounts for a systematic resummation of gradients in terms of trans-series, including non-analytical transient modes $\propto w^{-C_\tau/2C_\eta}e^{-3/2w}$, in the far-from-equilibrium medium~\cite{Basar:2015ava,Blaizot:2020gql}.  Therefore, following the strategy of effective field theory approach, fluctuating hydrodynamics on top of classical hydrodynamic attractor applies to noisy fluid out of equilibrium.
It is our purpose of the Letter, to investigate the effects of hydrodynamic fluctuations in a plasma out of equilibrium. 
Without loss of generality, we take \eq{eq:gcl} as the decay rate of the bare energy density, as the input for the next leading order analysis of fluctuating hydrodynamics.

{\it Hydrodynamic kinetic equation in the far-from-equilibrium regime.}---Bjorken symmetry is broken by fluctuations. 
In terms of Fourier modes of fluctuations of energy and momentum densities, 
\begin{subequations}
\begin{align}
\de(\tau, \vk)= \int d\zeta d^2 \vec x_\perp e^{i \vec k_\perp \cdot \vec x_\perp+i  \tau k_\zeta \zeta} \delta T^{\tau\tau}(\tau,\vec x_\perp, \zeta)\,,\\
g^i(\tau,\vk)=\int d\zeta d^2 \vec x_\perp e^{i \vec k_\perp \cdot \vec x_\perp+i  \tau k_\zeta \zeta} \delta T^{\tau i}(\tau,\vec x_\perp, \zeta)\,,
\end{align}
\end{subequations}
the equation $\partial_\mu \delta T^{\mu\nu}=-\partial_\mu S^{\mu\nu}$ leads to coupled stochastic differential equations for $\phi_a=(c_s \de, g^x, g^y,\tau g^\eta)$. Here $k_\zeta$ is dimensionful and conjugate to $\tau\zeta$.
These differential equations are equivalent to a hierarchy of equations for multi-point correlators~\cite{An:2020vri}. Especially, the equal-time two-point correlators, for which we define as $N_{ab}$ through
\be 
 \bra \phi_a(\tau,\vec k)\phi_b(\tau,-\vec k')\ket\equiv
 N_{ab}(\tau,\vec k) (2\pi)^3 \delta^3(\vec k - \vec k')\,,
\ee
satisfy effectively hydrodynamic kinetic equations~\cite{Akamatsu:2016llw}.  
Of course, when applies to out-of-equilibrium systems with large Kn, the background fluid evolution should be accounted by the hydrodynamic attractor.

To facilitate analyses, following Ref.~\cite{Akamatsu:2016llw}, it is convenient to rotate in the $\vec k$-space, 
which accordingly defines two longitudinal and two transverse modes $\phi_A$ and $A=(\pm,T_1,T_2)$. After the rotation, the hydrodynamic kinetic equation is dominated by the diagonal components. We therefore find formally
\be
\label{eq:eomRa}
\left(1+\frac{g(w)}{4}\right) \frac{\partial R_A}{\partial \ln w} = -\alpha_A w \tilde k^2(R_A -1 ) - \beta_A(w) R_A,
\ee
where $R_{A}\equiv \tau N_{AA}/T(e+P)$ is the normalized correlator and $\tilde k^2=\tau_\pi^2 (|\vec k_\perp|^2+k_\zeta^2)$. The coefficients are
\begin{align}
\alpha_{\pm} = \frac{4}{3 C_\tau}\,,\quad
\alpha_{T_1} = \alpha_{T_2} = \frac{2}{C_\tau}\,,
\end{align}
and
\begin{align}
\beta_{\pm}&= 1+ \frac{5}{4} g(w) +c_s^2 + \cos^2 \theta_k \,,\cr
\beta_{T_1}&=  1+ \frac{5}{4} g(w)\,, \quad
\beta_{T_2} =1 + \frac{5}{4}g(w)+2\sin^2\theta_k\,,
\end{align}
where $\cos\theta_k=k_\zeta/|\vec k|$.
The left hand side of \eq{eq:eomRa} represents time derivative on top of the background attractor. The first term ($\propto \alpha_A$) and the second term ($\propto \beta_A$) on the right hand side play the role of collision and longitudinal expansion, respectively. System evolution towards equilibrium relies then on these two competing effects.
Once \eq{eq:eomRa} is solved, the equal-time two-point correlators $N_{aa}$ in the original basis can be obtained respectively via an inverse rotation.

\begin{figure}
\begin{center}
\includegraphics[width=0.45\textwidth] {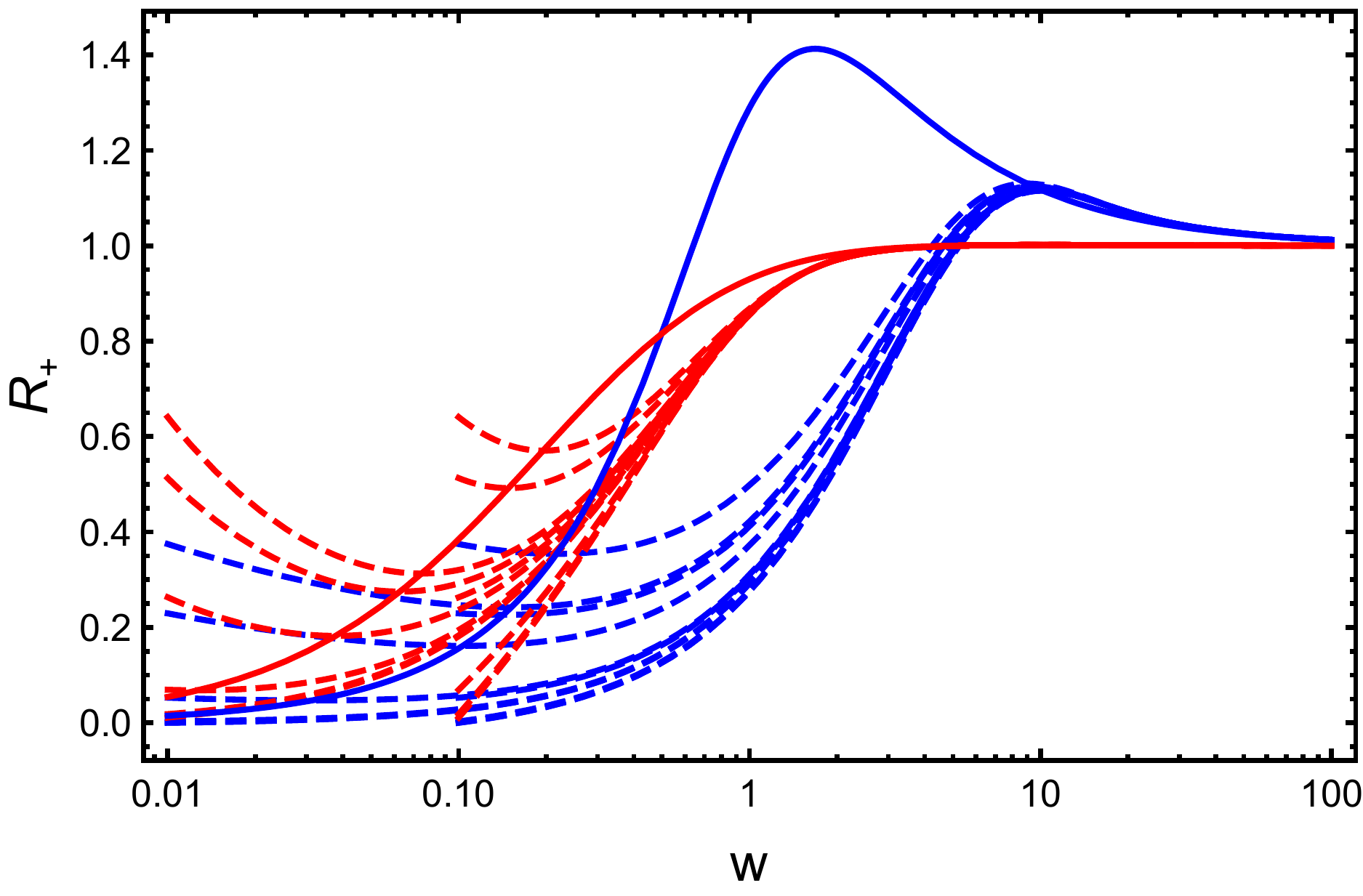}
\caption{
\label{fig:2pAtt}
Evolution of $R_+$ with respect to various initial conditions. Dashed blue and dashed red lines are obtained with $\tilde k=1,3$, and $\cos\theta_k=0.1,0.5$, respectively. Solid lines are the slow-roll approximation of the corresponding attractor, $R_+^{\rm slow-roll}=\alpha_+ w \tilde k^2/(\alpha_+ w \tilde k^2 + \beta_+)$.
}
\end{center}
\end{figure}

Hydrodynamic kinetic equation applies between separated scales: $c_s^{-1} \nabla\sim (c_s\tau)^{-1} \ll k \ll\mfp^{-1}$~\cite{Akamatsu:2016llw}. In the large $k$ limit, or more precisely when $w \tilde k^2\gg1$ according to \eq{eq:eomRa}, the two-point correlators approach 
$T(e+P)/\tau$, which can be understood as the ``equilibrium'' expectation in an out-of-equilibrium system defined according to background attractor. One thereby introduces a critical scale $k_*=\sqrt{w}/\tau$, above which the two-point correlators are well captured by the background attractor. In the out-of-equilibrium system, the separated scales, as well as $k_*$, are time dependent. In particular, because these scales merge around $w\approx 1$, a reliable description from \eq{eq:eomRa} for the out of equilibrium plasma should only apply when $w\in [1,+\infty)$.

As a consequence of fixed points, 
\eq{eq:eomRa} itself possesses attractor solutions. To see this, we first notice that $R_A=0$ is a fixed point solution in the small $w\tilde k^2$ extreme, which is stable only when $\beta_A>0$.  Nevertheless, the stability of this fixed point 
does not affect the two-point correlators at late times. In the large $w\tilde k^2$ extreme, \eq{eq:eomRa} allows for solution in terms of a double expansion
\be
\label{eq:Raexp}
R_A(w,w\tilde k^2) = 1 - \frac{\beta_A(w)}{\alpha_A w \tilde k^2} + \ldots = \sum_{n,m}\frac{F_{n,m}^{(A)}}{w^n (w \tilde k^2)^m}\,.
\ee
Note that the correlator depends explicitly on $w\tilde k^2$, as being dictated by \eq{eq:eomRa}.
The expansion in $1/w$ is the hydrodynamic gradient expansion of the two-point correlators, which is asymptotic. The expansion in $1/\tilde k^2$ is asymptotic as well, which, however, differs from the usual hydrodynamic gradient expansion in wavenumbers~\cite{Grozdanov:2019kge}. In the large-$w \tilde k^2$ extreme, the solution is well captured by the first several terms in the expansion. In analogy to the hydrodynamic fixed point in classical hydrodynamics 
at large $w$, the large-$w \tilde k^2$ behavior represents a stable hydrodynamic fixed point in the two-point correlators.

In \fig{fig:2pAtt}, for illustrative purposes, the evolution of $R_+$ is shown with two sets of $\tilde k$ and $\cos\theta_k$ values. 
Irrespective of initial conditions, $R_+$ tends to follow universal curves at late times (dashed lines), which is exactly the feature that one expects in an attractor solution. The universal curves stand for the attractors, which with slow-roll approximation ($\partial_w R_A=0$), can be approximated as $R_A^{\rm slow-roll} = \alpha_A w \tilde k^2/(\alpha_A w\tilde k^2 + \beta_A)$ 
(solid lines in \fig{fig:2pAtt}). Similar behavior can be found in other modes as well.

{\it Renormalization in far-from-equilibrium noisy fluids.}---The resulted equal-time two-point correlators $N_{aa}$ 
suffice to determine thermal corrections to the averaged energy-momentum tensor out of equilibrium. With respect to an integral in $\vec k$-space, the thermal corrections can be classified as a cut-off dependent correction $T^{\mu\nu}_\Lambda$ and a long-time tail correction $\Delta T^{\mu\nu}$. For instance, the $\tau\tau$-component,
\begin{align}
\label{eq:ttcor}
\bra T^{\tau\tau}\ket - T^{\tau\tau}_{\rm cl} & = 
\frac{1}{e+P}\int \frac{d^3 \vec k}{(2\pi)^3}\sum_{a=\vec x_\perp, \zeta} N_{aa}(w, w\tilde k^2)  \nonumber\\
&=   T^{\tau\tau}_\Lambda + \Delta T^{\tau\tau}\,,
\end{align}
where the explicit dependence on $w\tilde k^2$ is rooted in \eq{eq:eomRa}. 
As indicated in \eq{eq:Raexp}, the integral contains a cubic and a linear divergent pieces, which can be regulated by introducing a cut-off scale $\Lambda$. These regulated integrals then give rise to the cut-off dependent corrections. For $T^{\tau\tau}_\Lambda$ and $\tau^2 T^{\zeta\zeta}_\Lambda$, one finds,
\begin{subequations}\label{eq:correction}
\begin{align}
T^{\tau\tau}_\Lambda&= \frac{T\Lambda^3}{2\pi^2}- \frac{\Lambda T^3}{4\pi^2}\frac{C_\tau}{(C_\tau C_\eta)^2}
\frac{35}{8w}
\left(\frac{4}{3}+g(w)\right)\,,\\
\tau^2 T^{\zeta\zeta}_\Lambda 
& =\frac{T\Lambda^3}{6\pi^2}- \frac{\Lambda T^3}{4\pi^2}\frac{C_\tau}{(C_\tau C_\eta)^2}\frac{1}{w}
\left(\frac{27}{10}+\frac{35}{24} g(w)\right)\,.
\end{align}
\end{subequations}
In the limit $w\gg1$, these cut-off dependent corrections can be absorbed into the energy-momentum tensor, so that
energy density, pressure ($O(w^0)$)~\cite{Akamatsu:2016llw} and shear viscosity ($O(w^{-1})$) ~\cite{PhysRevD.84.025006} get renormalized, respectively. 
Note that since the cut-off dependent correction at $1/w$ is negative ($\propto -\Lambda T^3$), the resulted renormalized shear viscosity is actually enhanced.
In the far from equilibrium regime, with respect to the trans-series expansion in $g(w)$, higher order transport coefficients ($O(w^{-2}$) and beyond) are renormalized as well. 

Renormalization in fluctuating hydrodynamics reflects the fact that hydrodynamic fluctuations stay in equilibrium above the critical scale. 
In the out-of-equilibrium medium with large Kn, 
the cut-off scale can be taken according to $k_*\ll \Lambda \ll \mfp^{-1}$. In practice, provided the information of the physically measured quantities at a certain scale in the expanding system, $\Lambda$ is $w$ dependent and well constrained. 

{\it Out-of-equilibrium long time tails.}---After renormalization, a finite piece in the thermal corrections remains. As shown in \eq{eq:ttcor}, the explicit dependence on $w\tilde k^2$ implies an overall factor $w^{-3/2}$ in the finite integral, which leads to the non-analytical structure in the well-known long-time tails,
\begin{subequations}
\label{eq:ltt}
\begin{align}
\frac{\Delta T^{\tau\tau}}{e} &=\frac{w^{-3/2}}{C_e(C_\tau C_\eta)^3}\sum_{n=0} \frac{f_n^{\tau\tau}}{w^n} \,,\\
\frac{\tau^2\Delta T^{\zeta\zeta}}{e} &=  \frac{w^{-3/2}}{C_e(C_\tau C_\eta)^3}\sum_{n=0} \frac{f_n^{\zeta\zeta}}{w^n} \,.
\end{align}
\end{subequations}
Note that the long-time tails are cut-off independent.  The coefficients $f_n$ can be solved in principle by a summation of $F_{n,m}$ in \eq{eq:Raexp}.
Via a polynomial fit with respect to the numerical solutions of \eq{eq:eomRa}, we are allowed to identify $f_0^{\tau\tau}=0.45\pm0.1$, while $f_0^{\zeta\zeta}=0.15032\pm0.00002$ and $f_1^{\zeta\zeta}=-0.53\pm0.05$.

{\it Renormalized attractor. ---}
In fluctuating hydrodynamics, the effective out-of-equilibrium system evolution should be monitored by the thermal averaged components in the energy-momentum tensor. In particular, the renormalized ratio,
\begin{align}
\label{eq:renAtt}
\frac{\bra \tau^2 T^{\zeta\zeta}\ket}{\bra T^{\tau\tau}\ket}
& =\frac{\tau^2 T_{\rm cl}^{\zeta\zeta}+\tau^2 T^{\zeta\zeta}_\Lambda + \tau^2 \Delta T^{\zeta\zeta}}{T^{\tau\tau}_{\rm cl} + T^{\tau\tau}_\Lambda + \Delta T^{\tau\tau}}
\nonumber \\
& = [-1-g(w)] \Big(  1 + \frac{3\tau^2 T^{\zeta\zeta}_\Lambda}{e} +  \frac{3\tau^2 \Delta T^{\zeta\zeta}}{e} \nonumber\\
&\qquad\qquad\quad\qquad - \frac{T^{\tau}_\Lambda}{e} - \frac{\Delta T^{\tau\tau}}{e}  + \ldots \Big)\nonumber \\
& \equiv [-1-g(w)] Z^{-1}_{\rm att}(w)\,,
\end{align}
captures the observed system isotropization in the presence of hydrodynamic fluctuations.
In \eq{eq:renAtt}, a multiplicative renormalization factor $Z_{\rm att}^{-1}$ is introduced, which contains expansion in $1/w$ from the cut-off dependent corrections, and non-analytical corrections starting from $w^{-3/2}$ from the long time tails.

With respect to a {\it noisy} gluonic plasma~\cite{Heller_2013}, 
with $C_\eta=1/4\pi$, $C_\tau =2(2-\ln2) $ and $C_e=16\pi^2/30$, the renormalized attractor is solved and shown as the colored band in \fig{fig:renAtt}. The upper and lower boundaries are determined according to the two extreme cut-off scales, $\Lambda\sim k_*$ and $\Lambda\sim \mfp^{-1}$, respectively.  In both cases, when $\Lambda$ is explicitly taken into account, both the cut-off dependent corrections and the long-time tails are constrained by an overall factor $1/C_e (C_\tau C_\eta)^3$. 
 This factor is roughly the inverse of the number degrees of freedom in an unit volume, in consistency to the physical expectation of quadratic couplings of hydrodynamic fluctuations~\cite{Akamatsu:2016llw}. 

When $w\gg 1$, the system is close to an ideal fluid, hence effects of hydrodynamic fluctuations are expected suppressed. Indeed, at large $w$, the renormalized attractor follows the trend of classical hydrodynamics, approaching 1/3 irrespective of hydrodynamic fluctuations.  However, a closer look reveals that the renormalized attractor is actually below the classical result. 
This feature is expected, since 
the effective shear viscosity, which quantifies the reduction from 1/3, 
is enhanced by the renormalization due to hydrodynamic fluctuations~\cite{Akamatsu:2016llw,PhysRevD.84.025006}. Moreover, taking into account the fact that the renormalized correction to shear viscosity is proportional to $\Lambda$, a larger reduction in the 
renormalized attractor is expected with respect to a larger cut-off scale, as manifested in \fig{fig:renAtt}.

Unlike the classical attractor, which increases monotonically from far from equilibrium towards ideal fluids, the renormalized attractor becomes non-monotonic in the far-from-equilibrium region, as shown in \fig{fig:renAtt}. This non-monotonic behavior qualitatively reflects the long time tail contribution to $Z^{-1}_{\rm att}$. More precisely, the leading order long time tails give rise to positive corrections in $Z_{\rm att}^{-1}$, which compensates the reduction from the renormalized shear viscosity at large $w$~\cite{PhysRevD.84.025006}, but dominates when the system is far from equilibrium.   
Parametrically, by comparing the leading order long-time tails and the cut-off corrections in  $Z^{-1}_{\rm att}$ in the $\zeta\zeta$ sector, the minimum point of the renormalized attractor can be found around $\sqrt{w}\sim 4\pi^2 f_0^{\zeta\zeta}/C_\tau$.

{\it Summary and discussion.}---With the help of hydrodynamic attractor, fluctuating hydrodynamics can be applied to far-from-equilibrium noisy systems. As an example, the hydrodynamic kinetic equation, which was developed previously for a noisy fluid close to equilibrium, can be generalized to far-from-equilibrium noisy plasmas.
In the far-from-equilibrium medium, although hydrodynamic fluctuations lead to qualitatively similar contributions from the coupled modes, higher order contributions, such as terms of order $O(1/w^2)$ and $O(w^{-5/2})$, and even some more complicated trans-series structures in $g(w)$, are involved. Moreover, backreaction of the hydrodynamic fluctuations are significant in the far-from-equilibrium region, which modifies the system evolution towards equilibrium. In particular,  due to long time tails, in the noisy plasma the effective isotropization is non-monotonic. 

The current analysis could be improved systematically by including more ingredients in a noisy fluid system, such as the nonlinear couplings of modes beyond the quadratic order and non-conformal corrections. With respect to realistic high-energy nuclear collisions, initial state fluctuations, which differ conceptually from hydrodynamic fluctuations, should be taken into account as well.

{\it Acknowledgments.}---
 L.~Y. is supported by the National Science Foundation of China through grant No.~11975079. L.~Y. thanks Yi Yin for helpful discussions.

\bibliography{references}

\begin{thebibliography}{38}
\bibitem [{ff:()}]{ff:footnote}%
  \BibitemOpen
  \href@noop {} {}\bibinfo {note} {{With a matrix convention
  $g_{\mu\nu}=(-,+,+,+)$, the projection operators are $\Delta^{\mu\nu}= u^\mu
  u^\nu + g^{\mu\nu}$ and
  $\Delta^{\mu\nu\alpha\beta}=\frac{1}{2}(\Delta^{\mu\alpha}\Delta^{\nu\beta}+\Delta^{\mu\beta}\Delta^{\nu\beta})-\frac{1}{3}\Delta^{\mu\nu}\Delta^{\alpha\beta}$}}\BibitemShut
  {NoStop}%
\end{thebibliography}

\end{document}